\documentclass[10pt,letterpaper,conference]{IEEEtran}

\usepackage{amssymb}
\usepackage{amsmath}
\usepackage{graphicx}
\usepackage{psfrag}
\usepackage{color}

\psfrag{Crate}{$r$}
\psfrag{Qrate}{$S$}
\psfrag{R}{$R$}
\psfrag{S}{$S$}

\DeclareMathOperator{\tr}{Tr}

\begin{document}

\def\supp{ {\rm{supp \,}}}
\def\dist{ {\rm{dist }}}
\def\dim{ {\rm{dim \,}}}
\def\oti{{\otimes}}
\def\bra#1{{\langle #1 |  }}
\def\lb{ \left[ }
\def\rb{ \right]  }
\def\tilde{\widetilde}
\def\bar{\overline}
\def\*{\star}
\def\({\left(}		\def\BL{\Bigr(}
\def\){\right)}		\def\BR{\Bigr)}
	\def\BBL{\lb}
	\def\BBR{\rb}

\def\1{{\mathbf{1} }}

\def\bb{{\bar{b} }}
\def\ab{{\bar{a} }}
\def\zb{{\bar{z} }}
\def\zbar{{\bar{z} }}
\def\inv#1{{1 \over #1}}
\def\half{{1 \over 2}}
\def\d{\partial}
\def\der#1{{\partial \over \partial #1}}
\def\dd#1#2{{\partial #1 \over \partial #2}}
\def\vev#1{\langle #1 \rangle}
\def\ket#1{ | #1 \rangle}
\def\rvac{\hbox{$\vert 0\rangle$}}
\def\lvac{\hbox{$\langle 0 \vert $}}
\def\2pi{\hbox{$2\pi i$}}
\def\e#1{{\rm e}^{^{\textstyle #1}}}
\def\grad#1{\,\nabla\!_{{#1}}\,}
\def\dsl{\raise.15ex\hbox{/}\kern-.57em\partial}
\def\Dsl{\,\raise.15ex\hbox{/}\mkern-.13.5mu D}
\def\b#1{\mathbf{#1}}
\newcommand{\proj}[1]{\ket{#1}\bra{#1}}
\def\braket#1#2{\langle #1 | #2 \rangle}
%
%
\def\th{\theta}		\def\Th{\Theta}
\def\ga{\gamma}		\def\Ga{\Gamma}
\def\be{\beta}
\def\al{\alpha}
\def\ep{\epsilon}
\def\vep{\varepsilon}
\def\la{\lambda}	\def\La{\Lambda}
\def\de{\delta}		\def\De{\Delta}
\def\om{\omega}		\def\Om{\Omega}
\def\sig{\sigma}	\def\Sig{\Sigma}
\def\vphi{\varphi}
%
%
\def\CA{{\cal A}}	\def\CB{{\cal B}}	\def\CC{{\cal C}}
\def\CD{{\cal D}}	\def\CE{{\cal E}}	\def\CF{{\cal F}}
\def\CG{{\cal G}}	\def\CH{{\cal H}}	\def\CI{{\cal J}}
\def\CJ{{\cal J}}	\def\CK{{\cal K}}	\def\CL{{\cal L}}

\def\CM{{\cal M}}	\def\CN{{\cal N}}	\def\CO{{\cal O}}
\def\CP{{\cal P}}	\def\CQ{{\cal Q}}	\def\CR{{\cal R}}
\def\CS{{\cal S}}	\def\CT{{\cal T}}	\def\CU{{\cal U}}
\def\CV{{\cal V}}	\def\CW{{\cal W}}	\def\CX{{\cal X}}
\def\CY{{\cal Y}}	\def\CZ{{\cal Z}}

\def\rvac{\hbox{$\vert 0\rangle$}}
\def\lvac{\hbox{$\langle 0 \vert $}}
\def\comm#1#2{ \BBL\ #1\ ,\ #2 \BBR }
\def\2pi{\hbox{$2\pi i$}}
\def\e#1{{\rm e}^{^{\textstyle #1}}}
\def\grad#1{\,\nabla\!_{{#1}}\,}
\def\dsl{\raise.15ex\hbox{/}\kern-.57em\partial}
\def\Dsl{\,\raise.15ex\hbox{/}\mkern-.13.5mu D}
\def\beq{\begin {equation}}
\def\eeq{\end {equation}}
\def\to{\rightarrow}

\newtheorem{lem}{Lemma}
\newtheorem{prop}{Proposition}
\newtheorem{theo}{Theorem}
\newtheorem{dfn}{Definition}
\newtheorem{cor}{Corollary}
\def\ie{IEEEeqnarray}
\def\bIE#1{\begin{\ie}{#1}}
\def\eIE{\end{\ie}}
\def\nn{\nonumber}
\def\diag{\mbox{diag}}

\def\argmax{\mbox{argmax}}
\definecolor{gray}{gray}{.9}
\def\com#1{\vspace{.1in}\fcolorbox{black}{gray}{\begin{minipage}{5.5in}#1\end{minipage}}\vspace{.1in}}

\def\2ra{2^{nR_a}}
\def\2rb{2^{nQ_a}}
\def\2qa{2^{nR_b}}
\def\2qb{2^{nQ_b}}
\def\h#1{\widehat{#1}}

\title{Capacity Theorems for Quantum \\ Multiple Access Channels}
\author{\authorblockN{Jon Yard}
\authorblockA{Department of Electrical Engineering \\
Stanford University\\
Stanford, California, USA \\
{\tt jtyard@cs.mcgill.ca}
}
\and
\authorblockN{Igor Devetak}
\authorblockA{Electrical Engineering Department\\
University of Southern California\\
Los Angeles, California, USA \\
{\tt devetak@usc.edu}
}
\and
\authorblockN{Patrick Hayden}
\authorblockA{Computer Science Department \\
McGill University\\
Montr\'eal, Quebec, CA \\
{\tt patrick@cs.mcgill.ca}
}
}
\maketitle
\begin{abstract}
We consider quantum channels with two senders and one receiver.  For an arbitrary such channel, we give multi-letter characterizations of two different two-dimensional capacity regions.  The first region characterizes the rates at which it is possible for one sender to send classical information while the other sends quantum information.  The second region gives the rates at which each sender can send quantum information.  We give an example of a channel for which each region has a single-letter description, concluding with a characterization of the rates at which each user can simultaneously send classical and quantum information.
\end{abstract}

\section{Introduction}
Suppose that two independent senders, Alice and Bob, have access to different parts of the input of a quantum channel with a single receiver Charlie.  By preparing physical systems at their respective inputs, Alice and Bob can affect the state of Charlie's received system.  We will analyze situations in which such a channel is used many times for the purpose of sending independent information from each of Alice and Bob to Charlie.  We allow for the simultaneous transmission of both classical and quantum information.  

\section{Background}
A quantum system with the label $A$ is supported on a Hilbert space $\CH_A$ of 
dimension $|A|$. We will use a superscript to indicate that a density matrix $\rho^A$ or a pure state $\ket{\phi}^A$ belongs to the corresponding set of states of $A$.  By a channel $\CN^{A\rightarrow B}$, we mean a trace preserving, completely positive linear map from density matrices on $A$ to density matrices on $B$.  We will often just call this a map.  The tensor product $AB$ of two systems has a Hilbert space $\CH_{AB}\equiv\CH_A\otimes\CH_B$, and we will write $A^n$ for the system with Hilbert space $\CH_{A^n} \equiv \CH_A^{\otimes n}$.  The density matrix of a pure state $\ket{\phi}$ will be written as $\phi \equiv \proj{\phi}$.  When we speak of a \emph{rate $Q$ maximally entangled state}, we mean a bipartite pure state of the form
\[\ket{\Phi} = \frac{1}{\sqrt{2^{nQ}}}
 \sum_{b\in 2^{nQ}} \ket{b}\ket{b}\]
where $n$ will always be apparent from the context.
For a distance measure between two states $\rho$ and $\sigma$, we use the \emph{fidelity} 
\[F(\rho,\sigma) = \left(\tr\sqrt{\sqrt{\rho}\,\sigma\sqrt{\rho}}\right)^2.\]

\subsection{Classical-quantum states and entropy}
Consider a collection of density matrices $\big\{\sigma^B_x\big\}_{x\in\CX}$ indexed by a finite set $\CX$.  If those states occur according to the probability distribution $p(x)$, we may speak of an \emph{ensemble} $\big\{p(x),\sigma^B_x\big\}$ of quantum states.  Classical and quantum probabilities can be treated in the same framework by considering the \emph{cq system} $XB$ whose state is specified by a block-diagonal joint density matrix 
Such a \emph{cq state} describes the 
classical and quantum aspects of the ensemble on the \emph{extended Hilbert space} $\CH_X\otimes\CH_B$ \cite{dcr}.  Information quantities evaluated on cq states play an important role in characterizing the capacity regions we will introduce in this paper.
We write
$H(B)_\sigma = H(\sigma^B) = -\tr \sigma^B\log\sigma^B$
for the von Neumann entropy of the density matrix associated with $B$,
where $\sigma^B = \tr_{X}\sigma$.
Note that $H(X)_\sigma$ is just the Shannon entropy of $X$.
For an arbitrary state $\rho^{AB}$, we define
$H(AB)_\rho$ analogously.  The subscripted state will be omitted when it is apparent from the context.

\subsection{Classical capacity and mutual information}
The \emph{classical capacity} $C(\CN)$ of a quantum channel 
$\CN^{A'\rightarrow C}$ from Alice to Charlie is the 
logarithm of the number of physical input states Alice can 
prepare, per use of the channel, so that Charlie can reliably 
distinguish them with arbitrarily low probability of error.  $C(\CN)$ can be characterized in terms of the \emph{mutual information}, which is defined as
\bIE{C}
I(X;B) \equiv H(X) + H(B) - H(XB) \nn
\eIE
and is otherwise known as the Holevo information
$\chi\big(\{p(x),\sigma^B_x\}\big)$
of the underlying ensemble.
Mutual information gives a regularized characterization of the classical capacity $C(\CN)$ of $\CN$ as \cite{sw2, holcap}
\[C(\CN) = \lim_{k\rightarrow \infty} \frac{1}{k}\max_{XA^k} I(X;C^k)\]
where the maximization is over all states of a cq system $XA^k$ for which  $|\CX|\leq \min\{|A'|,|C|\}^{2k}$.  The mutual information is evaluated with respect to the induced state $\CN^{\otimes k}(\sigma)$ on $XC^k$.  Note that throughout this paper, we will adhere to the convention that an arbitrary channel $\CN$ acts as the identity on any system which is not explicitly part of its domain.  
The units of $C(\CN)$ are \emph{bits per channel use}.
It is a pressing open question of quantum Shannon theory whether or not it suffices to consider $k=1$ when computing $C(\CN)$.  Equivalently, it is unknown if entangled preparations are required to approach $C(\CN)$.

The \emph{quantum capacity} $Q(\CN)$ of $\CN$ gives the ultimate capability of $\CN$ to convey quantum information.  $Q(\CN)$ arises as the logarithm of various quantities divided by the number of channel uses.  Some of these include
\begin{itemize}
\item the amount of entanglement that can be created by using $\CN$
(entanglement generation) \cite{dev}
\item the size of a maximally entangled state that can be transmitted over $\CN$ (entanglement transmission) \cite{sn96}
\item the size of a perfect quantum channel from Alice to Bob which can be simulated with arbitrary accuracy (strong subspace transmission)
\cite{qmac1}.
\end{itemize}
All have units of \emph{qubits per channel use}.  $Q(\CN)$ can be characterized as a regularized maximization of \emph{coherent information}.
Depending on the context,
coherent information \cite{sn96} will be expressed in one of two ways.
For a fixed joint state $\sigma^{AB}$, we write
$I_c(A\,\rangle B) \equiv H(B) - H(AB) = -H(A|B).$
Otherwise, if we are given a density matrix $\rho^{A'}$ and a channel $\CN^{A'\rightarrow B}$ which give rise to a joint state 
$(1^A\otimes \CN)(\Phi_\rho)$,
where $\ket{\Phi_\rho}^{AA'}$ is any purification of $\rho$, we will often use the notation
\[I_c(A\,\rangle B) = I_c(\rho,\CN) = H(\CN(\rho)) - H((1\otimes\CN)(\Phi_\rho)).\]  
It can be shown that this latter expression is independent of the particular purification $\ket{\Phi_\rho}$ that is chosen for $\rho$.
The regularized expression for $Q(\CN)$ can thus be written in either of two equivalent ways as  
\begin{eqnarray*}
Q(\CN) = \lim_{k\rightarrow \infty} \frac{1}{k}\max_{AA'^k}I_c(A\,\rangle C^k) 
= \lim_{k\rightarrow \infty} \frac{1}{k}\max_{\rho^{A'^k}}I_c(\rho,\CN^{\otimes k}).
\end{eqnarray*}
In the first expression, the maximization is over all bipartite pure states $\ket{\Psi}^{AA'^k}.$ $I_c(A\,\rangle C^k)$ is then evaluated for the resulting state $\CN^{\otimes k}(\Psi)$ on $AC^k.$ 

We further remark that when $I_c(A\,\rangle BX)$ is evaluated on 
the cq state $\omega^{XAB} = \sum_x p(x)\proj{x}^X\otimes\omega_x^{AB},$
it can be considered as a conditional, or expected,
coherent information, as
$I_c(A\,\rangle BX)_\omega = \sum_x p(x) I(A\,\rangle B)_{\omega_x}.$
A particular departure of this quantity from its classical analog, the conditional mutual information $I(X;Y|Z)$, is that the latter is only equal to $I(X;YZ)$ when $X$ and $Z$ are independent, whereas the former always allows either interpretation, provided the conditioning variable is classical.  

Conditional coherent information arises in another context; suppose that $\boldsymbol{\CN}^{A'\rightarrow XB}$ is a \emph{quantum instrument} \cite{davies}, meaning that $\boldsymbol{\CN}$ acts as 
$\boldsymbol{\CN}\colon\tau \mapsto \sum_x \proj{x}^X \otimes \CN_x(\tau).$
The completely positive maps $\{\CN_x\}$ are the \emph{components} of the instrument.  While the components are generally trace reducing maps, their sum $\CN = \sum_x \CN_x$ is always trace preserving.  It is not difficult to show that 
$I_c(\rho,\boldsymbol{\CN}) = I_c(A\,\rangle BX),$
where the latter quantity is evaluated on the state
$\sum_x p(x) \proj{x}^X \otimes (1^A \otimes \CN_x)(\Phi_\rho^{AA'}).$

For us, a \emph{quantum multiple access channel} $\CN^{A'B'\rightarrow C}$ will have two senders and a single receiver.  While many-sender generalizations of our theorems are readily obtainable, we focus on two senders for simplicity.  Winter \cite{wintermac} gave a single-letter characterization of the rates at which classical information can be sent over a multiple access channel with classical inputs and a quantum output as the convex hull of a union of pentagons, a form identical to that found by Ahlswede \cite{ahlswede} and Liao \cite{liao} for the classical multiple access channel. For an arbitrary quantum multiple access channel, his results can easily be shown to yield a characterization of its classical capacity region in terms of a regularized union of pentagons.  Below, we summarize results appearing in  \cite{qmac1,yardthesis}, regarding the capabilities of quantum multiple access channels for sending classical and quantum information at the same time, while also supplementing that material with a new additive example.
\section{Strong subspace transmission}
Assume that Alice and Bob are connected to Charlie by $n$ instances of a multiple access channel  $\CN^{A' B' \rightarrow C}$, where Alice and Bob respectively have control over the $A'^n$ and $B'^n$ inputs. 
We will describe a scenario in which Alice wishes to transmit classical information at a rate of $R_a$ bits per channel use, while simultaneously transmitting quantum information at a rate of $Q_a$ qubits per channel use.  At the same time, Bob will be transmitting classical and quantum information at rates of $R_b$ and $Q_b$ respectively.  Alice attempts to convey any one of $2^{nR_a}$ messages to Charlie, while Bob tries to send him one of $2^{nR_b}$ such messages.  We will also assume that the senders are presented with systems $\tilde{A}$ and $\tilde{B}$,
where $|\tilde{A}| = 2^{nQ_a}$ and $|\tilde{B}| = 2^{nQ_b}$  Each will be required to complete the following two-fold task. Firstly, they must individually transfer the quantum information embodied in $\tilde{A}$ and $\tilde{B}$ to their respective inputs $A'^n$ and $B'^n$ of the channels, in such a way that it is recoverable by Charlie at the receiver.  Second, they must simultaneously make Charlie aware of their independent messages $M_a$ and $M_b$. 
Alice and Bob will encode with maps 
from the cq systems holding their classical and quantum messages to their respective inputs of $\CN^{\otimes n}$, which we denote
$\CE_a^{M_a \tilde{A} \rightarrow A'^n}$ and 
$\CE_b^{M_b \tilde{B} \rightarrow B'^n}$. 
Charlie decodes with a quantum instrument 
$\boldsymbol{\CD}^{C^n\rightarrow\h{M}_a\h{M}_b\h{A}\h{B}}.$ 
The output systems are assumed to be of the same sizes and dimensions as their respective input systems.  
For the quantum systems, we assume that there are pre-agreed upon unitary correspondences id$_a^{\tilde{A} \rightarrow \h{A}}$ and id$_b^{ \tilde{B}\rightarrow \h{B}}$ between the degrees of freedom in the quantum systems presented to Alice and Bob which embody the quantum information they are presented with and the target systems in Charlie's laboratory to which that information should be transferred.  
The goal for quantum communication will be to, in the strongest sense, simulate the actions of these corresponding identity channels.  We similarly demand low error probability for each pair of classical messages.  Formally, $(\CE_a,\CE_b,\boldsymbol{\CD})$ will be said to comprise an $(R_a,R_b,Q_a,Q_b,n,\epsilon)$ \emph{strong
subspace transmission code} for the channel $\CN$ if for all $m_a \in 2^{nR_a}$, $m_b \in 2^{nR_b}$, $\ket{\Psi_1}^{A\tilde{A}}$, $\ket{\Psi_2}^{B\tilde{B}}$, where $A$ and $B$ are purifying systems of arbitrary dimensions,  
\[F\Big(\ket{m_a}^{\widehat{M}_a}\ket{m_b}^{\widehat{M}_b}
\ket{\Psi_1}^{A\widehat{A}}\ket{\Psi_2}^{B\widehat{B}},
\Omega_{m_am_b}\Big) 
\geq 1-\epsilon\]
where 
\bIE{lCr}
\Omega_{m_a m_b}^{\widehat{M}_a\widehat{M}_bA\h{A}B\h{B}} =  
\boldsymbol{\CD}\circ\CN^{\otimes n}
\Big(\CE_a(\proj{m_a}^{M_a}\otimes\Psi_1^{A\tilde{A}}) & & \nn \\
\hspace{1.5in}\otimes\,\CE_b(\proj{m_b}^{M_b}\otimes\Psi_2^{B\tilde{B}})
\Big). \nn
\eIE
We will say that a rate vector $(R_a,R_b,Q_a,Q_b)$ is \emph{achievable} 
if there exists a sequence of  $(R_a,R_b,Q_a,Q_b,n,\epsilon_n)$ strong subspace transmission codes with $\epsilon_n \rightarrow 0$.  The simultaneous capacity region $\CS(\CN)$ is then defined as the closure of the collection of achievable rates.  Setting various rate pairs equal to zero uncovers six two-dimensional rate regions.  Our first theorem characterizes the two shadows relevant to the situation where one user only sends classical information, while the other only sends quantum information.  The next theorem describes the rates at which each sender can send quantum information.

\section{Classical-quantum capacity region $\CC\CQ(\CN)$}
Suppose that Alice only wishes to send classical information at a rate of $R$ bits per channel use, while Bob will only send quantum mechanically at $Q$ qubits per use of the channel. The rate pairs $(R,Q)$ at which this is possible comprise a classical-quantum (cq) region $\CC\CQ(\CN)$ consisting of rate vectors in $\CS(\CN)$ of the form $(R,0,0,Q)$.  Our first theorem describes $\CC\CQ(\CN)$ as a regularized union of rectangles.
\begin{theo}
 $\CC\CQ(\CN)$ = the closure of the union of pairs of nonnegative rates 
 $(R,Q)$ satisfying
\bIE{CCl}
R  \,\,\,&\leq&\,\,\, I(X;C^k)_\omega/k \nn\\
Q  \,\,\,&\leq&\,\,\, I_c(B\,\rangle C^kX)_\omega/k\nn
\eIE 
for some $k$, some pure state ensemble $\{p_x,\ket{\phi_x}^{A'^k}\}$ 
and some bipartite pure state $\ket{\Psi}^{BB'^k}$ giving rise to the state
\bIE{C}
\omega^{XBC^k} = \sum_x p_x \proj{x}^X\otimes \CN^{\otimes k}(\phi_x\otimes\Psi)). \label{cqarise}
\eIE
Further, it is sufficient to consider ensembles for which the number of elements satisfies  
$|\CX| \leq \max\{|A'|,|C|\}^{2k}.$
\end{theo}

In \cite{qmac1, yardthesis,cwit}, we showed that this characterization of $\CC\CQ$ can be single-letterized for a certain quantum erasure multiple access channel over which Alice noiselessly sends a bit to Charlie, while Bob sends him a qubit which is erased whenever Alice sends 1.  In this case, $\CC\CQ$ is given by those cq rate pairs $(R,Q)$ satisfying $0\leq R \leq H(p)$ and $0\leq Q \leq 1-2p$ for some $0\leq p\leq 1$.  In Section~\ref{section:qubitflip}, we demonstrate that a single-letter description of $\CC\CQ$ is also obtained for a particular ``collective qubit-flip channel."

\subsection{On the proof of Theorem 1}
In \cite{qmac1} we prove the coding theorem and converse for the simpler tasks of entanglement transmission (ET) and entanglement generation (EG) respectively, as each task is shown to yield the same achievable rates.
For entanglement transmission, Bob is only required to transmit half of a maximally entangled state, while Alice is satisfied with obtaining a low average probability of error for her classical messages.  Entanglement generation weakens this further, by only asking that Bob be able to create entanglement with Charlie, rather than preserve preexisting entanglement.  These relaxed conditions for successful quantum communication are directly analogous to the average probability of error requirement in the classical theory.  
By recycling a negligible amount of shared common randomness which is generated using the channel, it is possible to strengthen these less powerful codes to meet the requirements of strong subspace transmission.

Let us briefly sketch the coding theorem, referring the reader to \cite{qmac1} for further details, as well as for the converse.  Fixing a state $\omega^{XB''C}$ of the form (\ref{cqarise}), we prove the achievability of the corner point 
$(R,Q) = \big(I(X;C)-\delta,I_c(B''\,\rangle CX)-\delta\big)$,
where $\delta> 0$ is arbitrary.  For a suitably long blocklength $n$, Alice can encode her classical information with a random HSW code \cite{sw2,holcap}
for the channel 
$\CN_a(\tau^{A'}) = \CN(\tau\otimes\tr_{AC}\omega)$,
obtaining an arbitrarily low average probability of error. 
Simultaneously, Bob uses a random entanglement transmission code $(\CE,\CD)$ \cite{dev} capable of transmitting a rate $Q$ maximally entangled state $\ket{\Phi}^{B\tilde{B}}$ arbitrarily well 
over the \emph{instrument channel} $\boldsymbol{\CN}^{A'\to XC}$, where 
$\boldsymbol{\CN}(\tau^{B'})=  \sum_x p(x)\proj{x}^X  \otimes \CN(\phi_x\otimes \tau)$.
The randomness in the codes ensures that these channels are seen by each sender.  To decode, Charlie first measures $C^n$, learning Alice's classical message and causing a negligible disturbance to the global state.  He then uses Alice's message to simulate the instrument channel $\boldsymbol{\CN}$ so that he can apply the decoder $\CD$ from Bob's quantum code.  These decoding steps constitute the required decoding instrument $\boldsymbol{\CD}$, which can be shown to perform almost as well as the individual single-user codes, on average.  
The existence of a deterministic code performing as least as well is then inferred.

\section{Quantum-quantum capacity region $\CQ(\CN)$}
The situation in which each sender only attempts to convey quantum infomation to Charlie is described by the quantum-quantum (qq) rate region $\CQ(\CN)$ which consists of rate vectors in $\CS(\CN)$ of the form $(0,0,Q_a,Q_b)$.  Our second theorem gives a characterization of $\CQ(\CN)$ as a regularized union of pentagons. 
\begin{theo} \label{theo:qq}
$\CQ(\CN)$ = the closure of the union of pairs of nonnegative rates $(Q_a,Q_b)$ satisfying
\bIE{rCC}
Q_a \,\,\,&\leq&\,\,\,I_c(A\,\rangle BC^k)_\omega/k \nn\\
Q_b \,\,\, &\leq& \,\,\,I_c(B\,\rangle AC^k)_\omega/k\nn \\
Q_a+Q_b \,\,\, &\leq& \,\,\,I_c(AB\,\rangle C^k)_\omega/k \nn
\eIE 
for some $k$ and some bipartite pure states $\ket{\Psi_1}^{AA'^k}$, $\ket{\Psi_2}^{BB'^k}$ giving rise to the state
\bIE{C}
\omega^{ABC^k} = \CN^{\otimes k}(\Psi_1\otimes\Psi_2). \label{qqarise}
\eIE
\end{theo}

\subsection{On the proof of Theorem 2}
Let us briefly describe the proof of achievability for Theorem~2, referring the reader to \cite{qmac1} for the converse and other details.  As with Theorem~1, we focus on the transmission of maximal entanglement.  Fixing a joint state $\omega^{A''B''C}$ of the form (\ref{qqarise}) (with $k=1$), we achieve the corner point 
$(Q_a,Q_b) = \big(I_c(A''\,\rangle C)_\omega - \delta, I_c(B''\,\rangle A''C) - \delta\big)_\omega$
for any $\delta > 0$, by constructing a decoder which uses Alice's quantum information as side information for decoding Bob's quantum information.
Setting $\rho_1^{A'} = \tr_A \Psi_1$ and $\rho_2^{B'} = \tr_B \Psi_2$, 
define channels $\CN_a^{A'\rightarrow C}$ and $\CN_b^{B'\rightarrow A''C}$ by $\CN_a(\tau) =\CN(\tau\otimes\rho_2)$ and 
$\CN_b(\tau) =\CN(\Psi_1\otimes\tau),$
observing that 
$I_c(\rho_1, \CN_a) = I_c(A''\,\rangle C)$ and 
$I_c(\rho_2,\CN_b) = I_c(B''\,\rangle A''C).$
For suitably large $n$, 
there are arbitrarily good random entanglement transmission codes, $(\CE_a,\CD_a)$ for $\CN_a$ and $(\CE_b,\CD_b)$ for $\CN_b$, which transmit the respective halves of the rate $Q_a$ and $Q_b$ maximally entangled states $\ket{\Phi_a}^{A\tilde{A}}$ and $\ket{\Phi_b}^{B\tilde{B}}$. Alice encodes with some deterministic member of the random ensemble $\CE'_a$ while Bob randomly encodes with $\CE_b$, producing a random global state $\CN^{\otimes n}(\CE'_a(\Phi_a)\otimes\CE_b(\Phi_b))$ on $ABC^n$. 
Next, Charlie uses $\CD_a$ and $\CD_b$ to perform a sequence of operations which ultimately define his decoding operation 
$\CD^{C^n\rightarrow \widehat{A}\widehat{B}}$.   

As the randomness in Bob's code ensures that the joint state on $AC^n$ is approximately $\CN_a^{\otimes n}\circ\CE'_a(\Phi_a),$ Charlie can decode Alice's information first.
In order to simultaneously protect Bob's information, Charlie utilizes an 
\emph{isometric extension} $\CU'^{C^n\rightarrow \h{A}F}$ 
of the deterministic decoder $\CD'_a$.   Afterwards, he removes the system $\h{A}$ and 
replaces it with the $\h{A}$ part of a particular locally prepared state $\ket{\varphi}^{A''^n\widehat{A}}$.  He then inverts the decoding with the inverse $\CU^{-1}$ isometric extension of Alice's fully randomized decoder, leaving the remaining systems on $BC^nA''^n$ approximately in the state
$\CN_b^{\otimes n}\circ\CE_b(\Phi_b)$. Charlie then decodes Bob's state with $\CD_b$, yielding high fidelity versions of the original states $\ket{\Phi_a}$ and $\ket{\Phi_b}$ on average.  The protocol is then derandomized.

\subsection{History of the result}
In an earlier draft of \cite{qmac1}, we gave a characterization of  $\CQ(\CN)$ as the closure of a regularized union of rectangles, defined by 
$0\leq Q_a \leq  I_c(A\,\rangle C^k)/k$ and $0 \leq Q_b \leq I_c(B\,\rangle C^k)/k$.
This solution was conjectured on the basis of a duality between classical Slepian-Wolf distributed source coding and classical 
multiple access channels (see e.g.\ \cite{coverthomas}), as well as on a no-go theorem for distributed data compression of so-called irreducible pure state ensembles \cite{adhw}.  After the earlier preprint was available, Andreas Winter announced recent progress \cite{merge} with Michal Horodecki and Jonathan Oppenheim on the quantum Slepian-Wolf problem, offering a characterization identical in functional form to the classical one, while also supplying an interpretation of negative rates and evading the no-go theorem.  Motivated by the earlier mentioned duality, he informed us that he could prove that the qq capacity region could also be characterized in direct analogy to the classical case.  Subsequently, we found that we could modify our previous coding theorem to achieve the new region.  The newer characterization behaves better under single-letterization, as the following ``collective bit-flip channel" example illustrates.

\section{Collective qubit-flip channel}
\label{section:qubitflip}
Consider a channel into which both Alice and Bob input a single qubit.  With probability $p$, Charlie receives the qubits without error; otherwise, they are received after undergoing a $180^\circ$ rotation about the $x-$axis of the Bloch sphere.  The action of the channel can be written as 
\[\CN_p(\rho^{A'B'}) = (1-p)\rho + p (\sig_x\otimes \sig_x)\rho (\sig_x\otimes \sig_x),\]
where $\sig_x$ is the Pauli spin flip matrix.
We first summarize the argument from \cite{qmac1} which shows that $\CQ(\CN_p)$ is given by a single-letter formula.
 
Consider the following state of the form (\ref{qqarise}):
\[\om^{ABC} = (1-p)\Psi_+^{AC_A}\otimes\Psi_+^{BC_B} 
+ p \Psi_-^{AC_A}\otimes\Psi_-^{BC_B}.\]
Here, we define the Bell states $\ket{\Psi_{\pm}} = (\ket{00} \pm \ket{11}/\sqrt{2}$, 
identifying $C\equiv C_AC_B$.  By evaluating the single-letter rate bounds on $\omega$, one sees that $I_c(AB\,\rangle C)_\om = 2-H(p)$ and that $I_c(A\,\rangle BC)_\om = I_c(B\,\rangle AC)_\om = 1$. 
Since $\CN_p$ is a \emph{generalized dephasing channel} \cite{ds}, its quantum capacity is additive and can be calculated to be $2-H(p)$.  Clearly this is an upper bound on the maximum sum rate over $\CN_p$.
Observing that each bound is as large as it can be, we conclude that $\CQ(\CN_p)$ is given by a pentagon of nonnegative rates $(Q_a,Q_b)$ satisfying 
$Q_a+Q_b\leq 2-H(p)$ and $Q_a,Q_b \leq 1$.  Note that since the capacity of a single qubit bit-flip channel is $1-H(p)$, we may interpret this example as illustrating that if Alice codes for such a channel, while Bob performs no coding whatsoever, Charlie can still correct errors to Bob's inputs.
It is worth mentioning that computer calculations reveal that the older rectangle description of $\CQ(\CN_p)$ is \emph{non-additive}, indicating that the pentagon characterization is in fact more accurate than the rectangle one
for this channel.

In fact, $\CC\CQ(\CN_p)$ is also single-letter and is given by the \emph{same formula} as $\CQ(\CN_p)$, replacing $Q_a$ by $R$ and $Q_b$ by $Q$.  There are two ways to see this.  For the first, observe that $\CC\CQ(\CN_p)$ is the convex hull of the two rectangles corresponding to the states 
\begin{eqnarray*}
\om_1^{XBC} &=&
\frac{1}{2}\Big[\proj{0}^X\otimes\CN_p(\phi_+^{A'}\otimes\Psi_+^{BB'}) \\ 
& & \,\,\,\, + \,\, \proj{1}^X\otimes\CN_p(\phi_-^{A'}\otimes\Psi^{BB'}_+)\Big] \\
\om_2^{XBC} &=& 
\frac{1}{2}\Big[\proj{0}^X\otimes\CN_p(\phi_0^{A'}\otimes\Psi^{BB'}_+)  \\ 
& & \,\,\,\,\,\! + \,\, \proj{1}^X\otimes\CN_p(\phi_1^{A'}\otimes\Psi^{BB'}_+)\Big]
\end{eqnarray*}
for which $\ket{\phi_0} = \ket{0},$ $\ket{\phi_1} = \ket{1},$ and 
$\ket{\phi_\pm} = (\ket{0} \pm \ket{1})/\sqrt{2}$.
Namely, $(I(X;C),I_c(B\,\rangle CX))_{\om_1} = (1,1-H(p))$, 
while $(I(X;C),I_c(B\,\rangle CX))_{\om_2} = (1-H(p),1)$.
That this is the capacity region follows for the same reasons as with $\CQ(\CN_p)$; the rates are as large as can be, and are thus saturated.
For codes designed using $\om_1$, Alice's inputs are unaffected by the channel, while the effective channel from Bob to Charlie is one with quantum capacity equal to $1-H(p)$.  Codes may be designed for $\om_2$, in which Alice codes for a binary symmetric channel with parameter $p$, while Bob performs no coding whatsoever.  Charlie first decodes Alice's classical information, thereby learning the error locations so that he can correct Bob's quantum information.

While the coding theorem for the region of Theorem~1 employs decodings which use classical side information for decoding quantum information, it is also possible to use the techniques from the proof of Theorem~2 to prove a coding theorem achieving $(I(X;BC) -\delta, I_c(B\,\rangle C)-\delta)$, using quantum side information to decode the classical information. This yields a new characterization of $\CC\CQ$ as a regularized union of pentagons.  Specifically, $\CC\CQ(\CN)$ can be written as those pairs of nonnegative cq rates $(R,Q)$ satisfying
\begin{eqnarray*}
R &\leq& I(X;BC^k)_\om/k \\
Q &\leq& I_c(B\,\rangle C^kX)_\om/k\\
R+Q &\leq& \big[I(X;C^k)_\om + I_c(B\,\rangle C^kX)_\om\big]/k \\ 
&=& \big[I(X;BC^k)_\om + I_c(B\,\rangle C^k)_\om\big]/k.
\end{eqnarray*}  
for some $k\geq 0$ and some state $\sig^{XBC^k}$ of the form (\ref{cqarise}).
Evaluating these rate bounds (with $k=1$) for the state $\om_2$ gives the second way of deriving the single-letter characterization of $\CC\CQ(\CN_p)$ argued above, as the new corner point is $(I(X;BC),I_c(B\,\rangle C))_{\om_2} = (1,1-H(p)).$  The corresponding code is one in which Alice performs no coding, (even though her inputs are subject to noise),
while Bob codes for the single-user qubit-flip channel.
Charlie decodes Bob's quantum information first, which is then used to correct the errors in Alice's classical inputs, since Alice's inputs are subject to noise.  However, the new coding theorem yields no advantage when applied to $\om_1$, as 
$(I(X;BC),I_c(B\,\rangle C))_{\om_1} = (1-H(p),1)$ implies that the corresponding pentagon is actually a rectangle. 

\section{A characterization of $\CS(\CN)$}
Finally, we characterize the four-dimensional simultaneous capacity region $\CS(\CN)$ which was defined in Section~III.
\begin{theo}
$\CS(\CN) =$ the closure of the union of vectors
of nonnegative rates $(R_a,R_b,Q_a,Q_b)$ satisfying
\bIE{rCl}
R_a \,\,\,&\leq&\,\,\, I(X;C^k|Y)_\omega/k \nn\\
R_b \,\,\,&\leq&\,\,\,  I(Y;C^k|X)_\omega/k \nn\\
R_a + R_b \,\,\,&\leq&\,\,\,  I(XY;C^k)_\omega/k \nn\\
Q_a \,\,\,&\leq&\,\,\,  I_c(A\,\rangle C^kBXY)_\omega/k \nn\\
Q_b \,\,\,&\leq&\,\,\,  I_c(B\,\rangle C^kAXY)_\omega/k \nn\\
Q_a + Q_b \,\,\,&\leq&\,\,\,  I_c(AB\,\rangle C^kXY)_\omega/k\nn
\eIE
for some integer $k\geq 1$ and some bipartite pure state ensembles
$\{p(x),\ket{\psi_x}^{AA'^n}\}$, $\{p(y),\ket{\phi_y}^{BB'^n}\}$ giving rise to
\[\omega^{XYABC^k} =  
 \sum_{x,y}p(x)p(y)\,\proj{x}^X\otimes\proj{y}^Y \otimes \omega_{xy}\]
where 
$\omega^{ABC^k}_{xy} = \CN(\psi_x^{AA'^k}\otimes\phi_y^{BB'^k}).$
Furthermore, it suffices to consider ensembles for which 
$|\CX| \leq \min\{{|A'|,|C|}\}^{2k}$ and $|\CY| \leq \min\{{|B'|,|C|}\}^{2k}$.
\end{theo}

This characterization of $\CS(\CN)$ generalizes a number of existing results in the literature.  Setting the quantum rates equal to zero yields a region which is the regularized optimization over input ensembles of the region given by Winter in \cite{wintermac} for  classical capacity of a classical-quantum multiple access channel.  By setting both of either Alice's or Bob's rates equal to zero, the result of Devetak and Shor \cite{ds} on the simultaneous classical-quantum capacity region of a single-user channel follows.  Two of the remaining three shadows are instances of our Theorem~1, while the final one gives Theorem~2.  We remark that the pentagon characterization of $\CC\CQ(\CN)$ does not follow as a corollary since, as we will see, all of the classical information is decoded first.

Briefly, achievability of $\CS(\CN)$ is obtained as follows.  Using techniques introduced in \cite{ds}, each sender ``shapes" their quantum information into HSW codewords.  Decoding is accomplished by first decoding all of the classical information as with $\CC\CQ(\CN)$, after which techniques from \cite{ds} as well as those used to achieve $\CQ(\CN)$ are utilized.  

\emph{Acknowlegements:}  JY would like to thank T.\ Cover for much useful input and feedback and Y.\ H.\ Kim for useful discussions regarding classical multiple access channels.  JY has been supported by the Army Research Office MURI under contract DAAD-19-99-1-0215, the Canadian Institute for Advanced Research, the National Science Foundation under grant CCR-0311633, and the Stanford Networking Research Center under grant 1059371-6-WAYTE.
PH is grateful for support from the Canadian Institute for Advanced Research, the Sherman Fairchild Foundation, the NSERC of Canada and the US National Science Foundation under grant EIA-0086038.


\begin{thebibliography}{50}
\bibitem{dcr} I. Devetak, A. Winter, ``Distilling common randomness from bipartite quantum states," {\tt quant-ph/0304196}, 2003.

\bibitem{dev} I. Devetak, ``The private classical information capacity and quantum information capacity of a quantum channel," {\tt quant-ph/0304127}.

\bibitem{sn96} B. Schumacher, M. A. Nielsen, ``Quantum data processing and error correction," \emph{Phys.\ Rev.\ A}, 54(4)2629, 1996.

\bibitem{qmac1} J. Yard, I. Devetak, P. Hayden, ``Capacity theorems for quantum multiple access channels -- classical-quantum and quantum-quantum capacity regions," 
{\tt quant-ph/0501045}. 

\bibitem{yardthesis} J. Yard, ``Simultaneous classical-quantum capacities of quantum multiple access channels," Ph.D. thesis, Electrical Engineering Dept.\ \!\!, Stanford University, {\tt quant-ph/0506050}, March 2005.

\bibitem{cwit} J. Yard, I. Devetak, P. Hayden, ``Sending classical and quantum information over quantum multiple access channels," \emph{Proc.\ Ninth Annual Canadian Workshop on Inform.\ Theory,} Montr\'eal, Canada, 2005.

\bibitem{davies} E. B. Davies, J. T. Lewis, ``An operational approach to quantum probability," Comm. Math. Phys., vol. 17, pp. 239-260, 1970. 

\bibitem{wintermac} A. Winter, ``The capacity of the quantum multiple access channel," \emph{IEEE Trans.\ Info.\ Th.,}
vol. IT-47, pp. 3059 - 3065, November 2001.

\bibitem{ahlswede} R. Ahlswede, ``Multi-way communication channels," Second Intern.\ Sympos.\ on Inf.\ Theory, Thakadsor, 1971, Publ.\ House of the Hungarian Adad.\ of Sciences, 23-52, 1973.

\bibitem{liao} Liao, ``Multiple access channels", Ph.D.\ dissertation, Dept.\ of Electrical Engineering, University of Hawaii, 1972.

\bibitem{bds} C. Bennett, D. DiVincenzo, J. Smolin, ``Capacities of quantum erasure channels," 
{\tt quant-ph/9701015}.

\bibitem{sw2} B. Schumacher, M. D. Westmoreland, ``Sending classical information via noisy quantum channels," \emph{Phys.\ Rev.\ A, vol.\ 56, no.\ 1, p.\ 131}.

\bibitem{holcap} A. S. Holevo, ``The capacity of the quantum channel with general input states," \emph{IEEE Trans.\ Info.\ Th.\ vol.\ 44, no.\ 1, p.\ 269}.

\bibitem{coverthomas} T. Cover, J. A. Thomas, ``Elements of information theory," John-Wiley \& Sons, Inc., 1991.

\bibitem{adhw} C. Ahn, P. Doherty, P. Hayden, A. Winter, ``On the distributed compression of quantum information," {\tt quant-ph/0403042}.

\bibitem{merge} M. Horodecki, J. Oppenheim, A. Winter, ``Partial quantum information," \emph{Nature}, vol.\ 436, no.\ 4, pp.\ 673--676, {\tt quant-ph/0505062}.

\bibitem{ds} I. Devetak, P. Shor, ``The capacity of a quantum channel for simultaneous transmission of classical and quantum information," \\ {\tt quant-ph/0311131}.


\end{thebibliography}
\end{document}